\documentclass[aps,prl,reprint,groupedaddress]{revtex4-1}

\usepackage{graphicx}

\begin{document}

\title{Thermalization and Sub-Poissonian Density Fluctuations\\
in a Degenerate Molecular Fermi Gas}
\author{William G. Tobias${}^*$}
\author{Kyle Matsuda${}^*$}
\author{Giacomo Valtolina}
\author{Luigi De Marco}
\author{Jun-Ru Li}
\author{Jun Ye}
\affiliation{JILA, National Institute of Standards and Technology and Department of Physics, University of Colorado, Boulder, Colorado 80309, USA}
\date{\today}
\begin{abstract}
    We observe thermalization in the production of a degenerate Fermi gas of polar ${}^{40}\text{K}{}^{87}\text{Rb}$ molecules. By measuring the atom--dimer elastic scattering cross section near the Feshbach resonance, we show that Feshbach molecules rapidly reach thermal equilibrium with both parent atomic species. Equilibrium is essentially maintained through coherent transfer to the ground state. Sub-Poissonian density fluctuations in Feshbach and ground-state molecules are measured, giving an independent characterization of degeneracy and directly probing the molecular Fermi--Dirac distribution.
\end{abstract}
\maketitle

Degenerate gases of polar molecules, which exhibit long-range, anisotropic dipole-dipole interactions, open new possibilities for engineering strongly-correlated quantum matter \cite{Micheli2006,Baranov2012,Hazzard2013,Yao2013,Syzranov2014,Baranov2002,Miyakawa2008,Veljic2018,Yan2013}. Heteronuclear molecules have been produced near quantum degeneracy by magnetoassociation of weakly-bound Feshbach molecules followed by coherent optical transfer to the rovibrational ground state \cite{Ni2008,Takekoshi2014,Molony2014,Park2015a,Guo2016,Rvachov2017,Seebelberg2018,Yang2019}. Recently, a degenerate Fermi gas of polar ${}^{40}\text{K}{}^{87}\text{Rb}$ molecules was realized using this method, starting from a deeply degenerate Bose--Fermi atomic mixture \cite{DeMarco2019}. The degenerate molecules were found to have momentum distributions consistent with thermal equilibrium and exhibited reduced chemical reactivity due to quantum statistics.

Reaching higher phase space density remains an outstanding challenge in ultracold molecule experiments. Multiple factors hinder efficient evaporation of ground-state molecules, including inelastic loss \cite{Ospelkaus2010,Mayle2013} and weak elastic interactions in the absence of an applied electric field \cite{Quemener2011}. Producing degenerate Feshbach molecules can thus be critically important for creating degenerate ground-state molecules. Feshbach molecule conditions may depend sensitively on atom--dimer thermalization during the molecule association process, which has not been studied in experiment.

Bosonic Feshbach molecules formed in Fermi--Fermi mixtures are observed to reach thermal equilibrium due to strong atom--dimer and dimer--dimer elastic interactions and fermionic suppression of inelastic processes \cite{Greiner2003,Regal2004,Jochim2003,Zwierlein2004,Strecker2003,Jag2014,Jag2016}. For heteronuclear molecules produced from Bose--Fermi mixtures, the situation is more complex. Inelastic boson--dimer collisions play a larger role \cite{Zirbel2008,Bloom2013,Johansen2017} and atom--dimer elastic scattering has not been previously measured.
Characterizing elastic and inelastic processes in these systems is essential for understanding thermalization dynamics and optimizing the production of a low-entropy sample.

In this Letter, we demonstrate that  ${}^{40}\text{K}{}^{87}\text{Rb}$ Feshbach molecules (KRb*) produced from a degenerate Bose--Fermi mixture rapidly come to thermal equilibrium, and that equilibrium is essentially maintained after coherent transfer to ground-state molecules (KRb). To quantify elastic processes during KRb* formation, we measure the magnitude of the atom--dimer scattering length for K--KRb* and Rb--KRb* collisions as a function of the magnetic bias field. We find that the molecular degeneracy saturates as a function of the magnetoassociation ramp rate, indicating that elastic collisions predominate over inelastic collisions and lead to thermalization.  As a direct probe of the state occupation of degenerate molecular samples, we additionally measure sub-Poissonian number fluctuations in KRb* and KRb, a technique previously used to characterize degeneracy and phase transitions in atomic gases \cite{Esteve2006,Gemelke2009,Sanner2010,Muller2010,Amico2018,Omran2015}. The momentum distribution in time-of-flight (TOF) expansion and the spatial profile of density fluctuations give consistent results for the molecular $T/T_{\text{F}}$, where $T_{\text{F}}$ is the Fermi temperature, validating the thermometry of the gas. 

We prepare an ultracold mixture of fermionic ${}^{40}\text{K}$ in the $|F, m_F\rangle = |9/2, -9/2\rangle$ hyperfine state and bosonic ${}^{87}\text{Rb}$ in the $|1,1\rangle$ state in a crossed optical dipole trap. The trap frequencies are $(\omega_{x}, \omega_{y},\omega_{z}) = 2\pi \times (60, 240, 60) $~Hz for K, and are scaled by factors of 0.72, 0.83, and 0.79 for Rb, KRb*, and KRb, respectively. The trap $y$-axis and the bias magnetic field $B$ are aligned in the direction of gravity. Feshbach molecules are produced by ramping $B$ through the broad interspecies resonance at $B_0 = 546.62$ G (3.04 G width) \cite{Klempt2008}, and can be subsequently transferred to the ground state using stimulated Raman adiabatic passage (STIRAP). Initial atom conditions for optimal molecule production are $5 \times 10^5$ K at $T/T_{\text{F}} = 0.1$ and $6 \times 10^4$ Rb at $T/T_{\text{c}} = 0.5$, resulting in $3\times 10^4$ KRb molecules at $T/T_{\text{F}} = 0.3$ \cite{DeMarco2019}.

The interplay of atom--atom and atom--dimer elastic and inelastic processes, which depend on the detuning from the Feshbach resonance, leads to a complicated evolution of the K--Rb--KRb* mixture during the Feshbach ramp. Inelastic processes in this system have been previously characterized experimentally \cite{Zirbel2008,Bloom2013}. Near the Feshbach resonance, free K and Rb atoms are indistinguishable from the weakly-bound molecular constituents, leading to fermionic suppression of inelastic collisions of KRb* with K and bosonic enhancement of those with Rb. In order to minimize inelastic Rb--KRb* losses, the initial Rb number is chosen so that Rb is no longer condensed after molecule production. After forming molecules, the peak density of K is approximately 10 times larger than Rb, so thermalization of KRb* is expected to occur predominantly through collisions with K. Despite the low Rb density, we expect a small number of Rb-KRb* elastic collisions to occur during the Feshbach ramp; measurements of the Rb-KRb* scattering length are included in the Supplementary Material \cite{SI}.

Here, we extract the elastic cross section for K--KRb* scattering as a function of $B$ by measuring the damping of KRb* center-of-mass oscillations due to collisions with K. The damping rate is proportional to the elastic collision rate $\Gamma = \overline n \sigma v_{\text{rel}}$, where $\overline n = \left(\frac{1}{N_{\text{K}}} + \frac{1}{N_{\text{KRb*}}}\right) \int n_{\text{K}} n_{\text{KRb*}} \; d^3 x$ is the overlap density between the two species, $n_{\text{K}}$ ($n_{\text{KRb*}}$) is the K (KRb*) density distribution, $\sigma$ is the K--KRb* elastic cross section, and $v_{\text{rel}} = \left[\frac{8 k_{\text{B}}}{\pi} \left(\frac{T_{\text{K}}}{m_{\text{K}}} + \frac{T_{\text{KRb*}}}{m_{\text{KRb}}} \right)\right]^{1/2}$ is the average relative velocity between the two species \cite{Maddaloni2000,Gensemer2001,Ferrari2002,Ferlaino2003,Naik2011}. We assume $s$-wave atom--dimer collisions, so that $\sigma = 4\pi a_{\text{ad}}^2 / (1 + k_{\text{th}}^2 a_{\text{ad}}^2)$, where $a_{\text{ad}}$ is the K--KRb* atom--dimer $s$-wave scattering length, $k_{\text{th}} = \sqrt{2 \mu k_B T/ \hbar^2}$ is the thermal collision wavevector, and $\mu$ is the K--KRb* reduced mass \cite{Chin2010}. A universal prediction gives $a_{\text{ad}} = 1.09 a$ near the Feshbach resonance for the mass ratio $m_{\text{K}} / m_{\text{Rb}} = 0.46$, where $a$ is the K--Rb scattering length \cite{Petrov2003}.

\begin{figure}[ht]
    \includegraphics[width=0.5\textwidth]{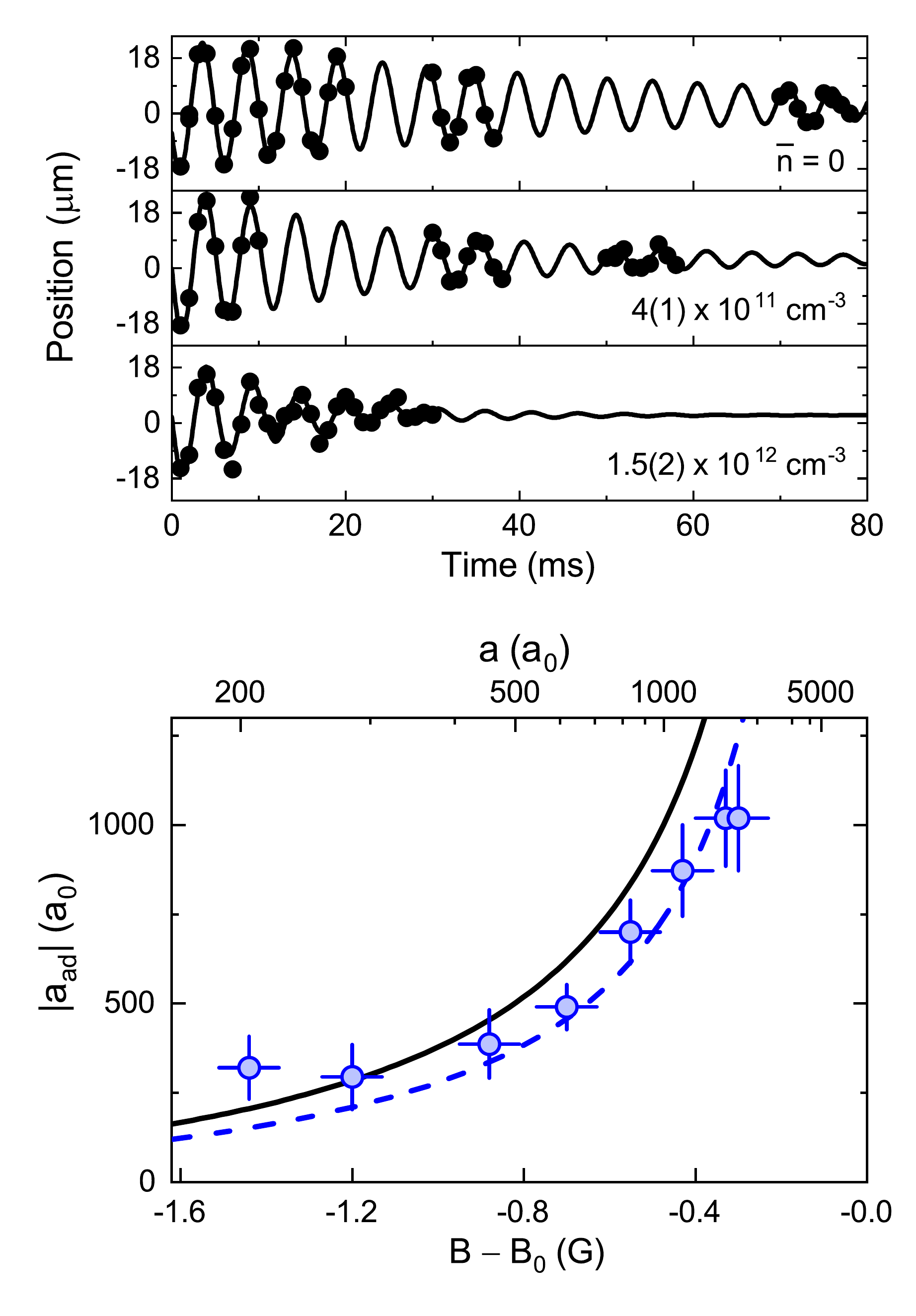}
    \caption{Upper panel: example oscillations of KRb* at $B = 546.1$ G, varying the overlap density $\overline n$. The values of $|a_{\text{ad}}| = 684(110)a_0$ and $648(88)a_0$ extracted from the middle and bottom data, respectively, are in good agreement. Lower panel: $|a_{\text{ad}}|$ vs $B-B_0$. Dashed line is a fit to the experimental data; see main text. For comparison, the K--Rb scattering length $a(B)$ is shown (solid line). The overlap density for these measurements is $\overline n = 1.0(1) \times 10^{12} \text{ cm}^{-3}$. Vertical error bars denote the standard error; horizontal error bars reflect a small settling of \textit{B} during the hold time. }

    \label{fig:a_ad}
\end{figure}

To perform the measurement, we first produce ground-state KRb at $B = 545.5$ G and remove all of the Rb atoms and a fraction of the K atoms using a combination of microwave pulses and light resonant with the atomic transition.
A second STIRAP sequence transfers the molecules back to the Feshbach state, producing a sample of $2 \times 10^4$ KRb* at $T = 300$ nK and $1.5 \times 10^5$ K at $T = 600$ nK. The timing of the two STIRAP pulses is chosen such that the photon recoil selectively excites a center-of-mass oscillation of KRb* \cite{SI}. Next, we ramp to the target $B$ in 0.5 ms, and after a variable hold time image the position of KRb* after 6 ms TOF. Due to the large number imbalance between K and KRb*, no induced motion of K is observed. By fitting the decay of the KRb* oscillations, corrected by a small background damping rate due to trap anharmonicity, we obtain a measurement of the collision rate $\Gamma$ and extract $|a_{\text{ad}}|$.
The upper panel of Fig. \ref{fig:a_ad} shows the KRb* oscillations at a field of $B = 546.1$ G with varying overlap density $\overline n$. Over nearly a factor of four in $\overline n$, we extract consistent values of $|a_{\text{ad}}|$.
Since K is initially much hotter than KRb*, KRb* rapidly heats to the temperature of K. Measuring the timescale of this heating gives a consistent measurement of the atom--dimer cross section \cite{SI}. 

The results are summarized in the lower panel of Fig. \ref{fig:a_ad}, which shows the extracted $|a_{\text{ad}}|$ as a function of $B - B_0$. We perform measurements in the region $a < 2000 a_0$ ($B - B_0 < -0.3$~G), where the average collision energy is lower than the binding energy. The measured $|a_{\text{ad}}|$ as a function of $B$ is fit with a single parameter $c$, which accounts for a scaling $|a_{\text{ad}}| = ca$. The best fit gives $c = 0.74(5)$. Due to the high collision energy and relatively low atom--atom scattering lengths considered here, we do not necessarily expect that the universal prediction $a_{\text{ad}} = 1.09 a$ holds \cite{Petrov2003}. Nonetheless, we measure a large atom--dimer scattering length whose magnitude increases near the resonance. An estimate using our typical atomic and molecular densities and the measured $|a_{\text{ad}}|$ suggests more than 6 elastic collisions per molecule occur during a 5 ms Feshbach ramp from 555 G to 545.5 G (1.9 G/ms ramp rate), enabling thermalization \cite{SI}.

\begin{figure}
\centering
\hspace{-2em}
\includegraphics[width=0.48\textwidth]{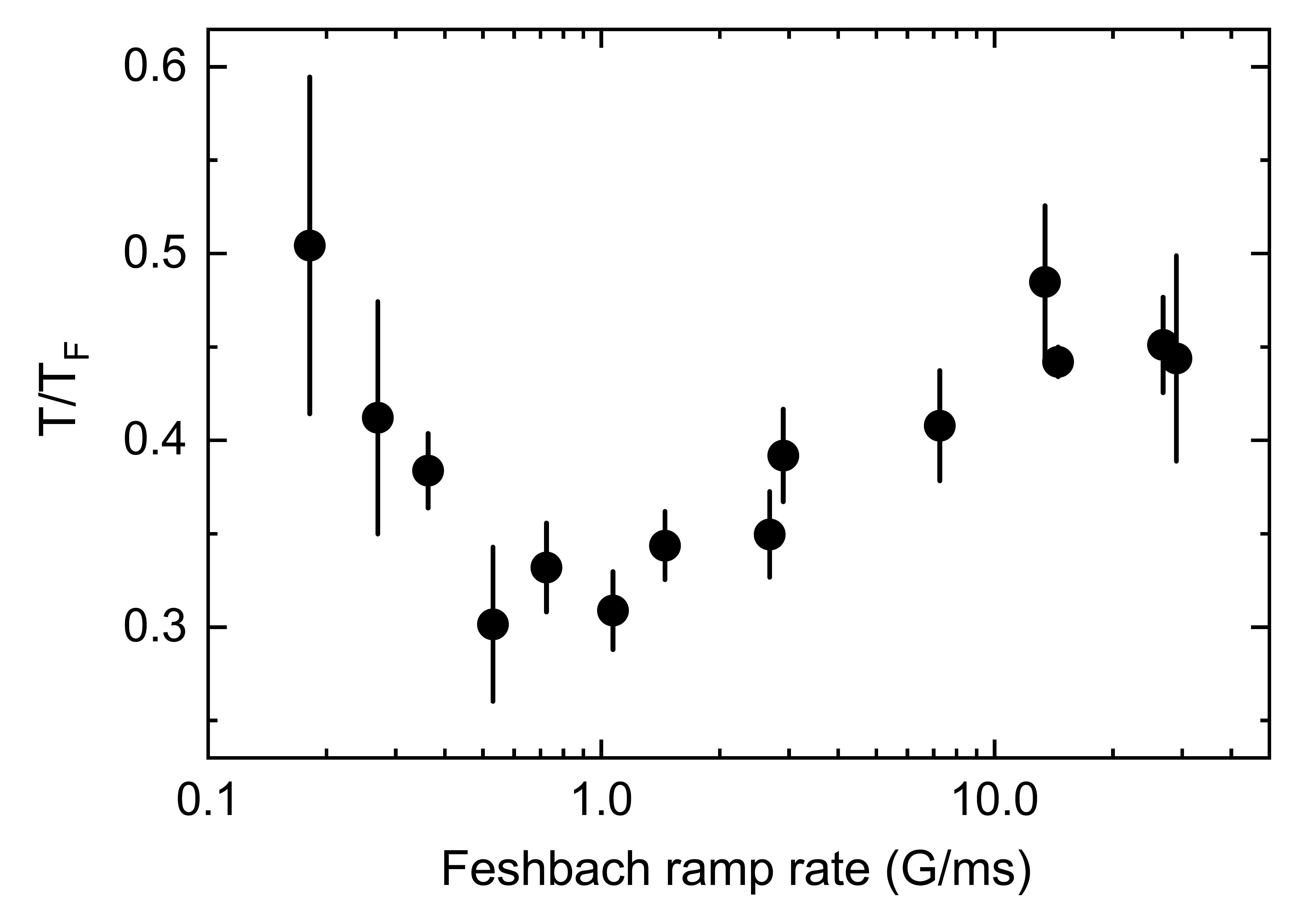}
\caption{KRb $T/T_{\text{F}}$ vs Feshbach ramp rate. Error bars denote the standard error. The most degenerate molecules are created with intermediate ramp rates of 0.5--3 G/ms.}
\label{fig:ramprate}
\end{figure}

Varying the Feshbach ramp rate provides an additional method for probing the balance between elastic and inelastic scattering rates \cite{Greiner2003}. In the molecule creation process, there is a competition between thermalization, which favors slower ramp rates, and inelastic losses, which are minimized with faster ramp rates. Figure \ref{fig:ramprate} shows the $T/T_{\text{F}}$ of the KRb cloud, measured by fitting the shape of the cloud after TOF expansion, as a function of the $B$ ramp rate. For intermediate ramp rates of 0.5--3 G/ms, $T/T_{\text{F}}$ reaches a minimum at 0.3. At very slow ramp rates ($<0.5$ G/ms), we observe substantial loss from inelastic processes, resulting in a sharp increase in the molecular $T/T_{\text{F}}$. We also observe a gradual rise in $T/T_{\text{F}}$ as the ramp time becomes much shorter than the trap oscillation period, while the molecule number remains constant, suggesting that thermalization is hindered for fast ramp speeds.

To confirm that the molecules are in thermal equilibrium, we measure the number fluctuations in the gas as an independent probe of $T/T_{\text{F}}$. Within a subvolume of a classical gas at any temperature, fluctuations are Poissonian, meaning the particle number variance over many experimental repetitions is equal to the mean: $\sigma_N^2/\overline{N}=1$. In the case of a Fermi gas with $T/T_{\text{F}}\ll1$, where nearly all states below the Fermi energy are singly occupied, the peak variance is suppressed below the mean by the factor $(3/2)T/T_{\text{F}}$ \cite{Sanner2010}. The quantity $\sigma_N^2/\overline{N}$ is therefore directly related to the degeneracy of the sample and provides a local measurement of state occupation.

The $T/T_{\text{F}}$ of molecules has so far been measured by fitting the momentum distribution after free expansion. The result is in good agreement with the ratio of the measured temperature and the $T_{\text{F}}$ calculated from the molecule number and trap frequencies. One important consideration for thermometry of ground-state molecules is the effect of STIRAP, which uniformly introduces a small number of holes in the KRb state distribution while preserving the shape of the expanded cloud. Here, we validate expansion thermometry by measuring local number fluctuations in degenerate molecular gases, and show that STIRAP has only a small effect on the molecular state occupation even for the lowest temperatures achieved.

We perform measurements of number fluctuations on K, KRb*, and KRb after 6 ms of free expansion \cite{SI}.
For measurements on KRb, two successive STIRAP sequences are used: the first converts KRb* to KRb and the second converts KRb to KRb* for imaging prior to TOF expansion. The total duration of the two STIRAP sequences is 270 $\mu$s, short enough that inelastic losses between KRb and the remaining atoms are negligible \cite{Ospelkaus2010}. In order to accurately count the molecule number, we adiabatically dissociate KRb* during expansion with a 2.5 G/ms ramp of \textit{B} before imaging \cite{Mukaiyama2004}. Sub-Poissonian number fluctuations have not been previously observed in molecules, making  characterization of the imaging system and of the data analysis procedure essential. We use measurements on degenerate and non-degenerate K atoms to benchmark the experimental methods against previous studies on atomic Fermi gases \cite{Sanner2010,Muller2010}.

\begin{figure}[hb]
\includegraphics[scale=0.35]{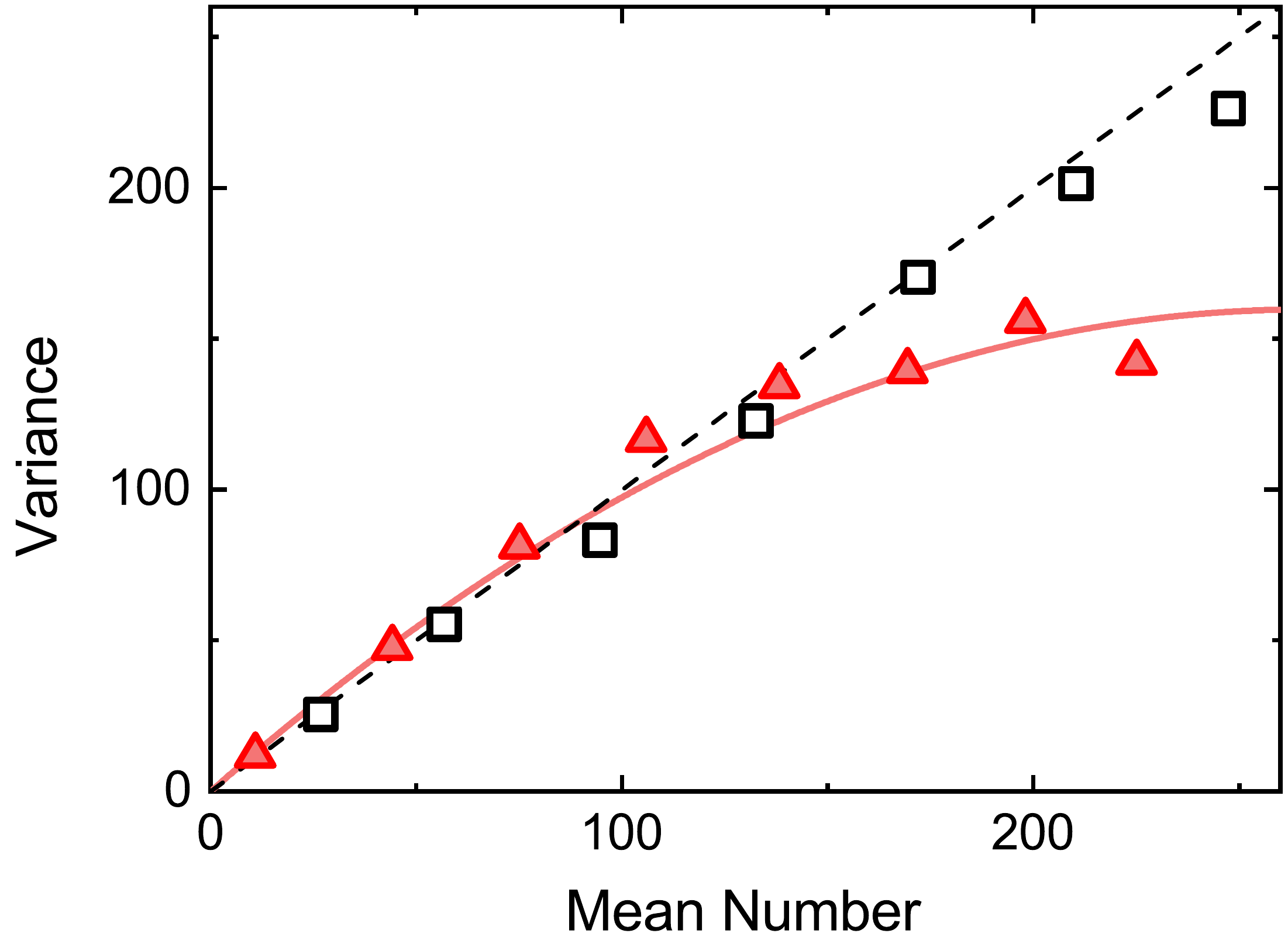}
\caption{Variance vs. mean number for non-degenerate K (open squares) and degenerate KRb (solid triangles), averaged over bins with similar mean number. The dashed line indicates equal mean and variance, the expected result for classical particles, and the solid curve is a guide to the eye.}
\label{fig:krb}
\end{figure}
\begin{figure*}
\includegraphics[scale=0.53]{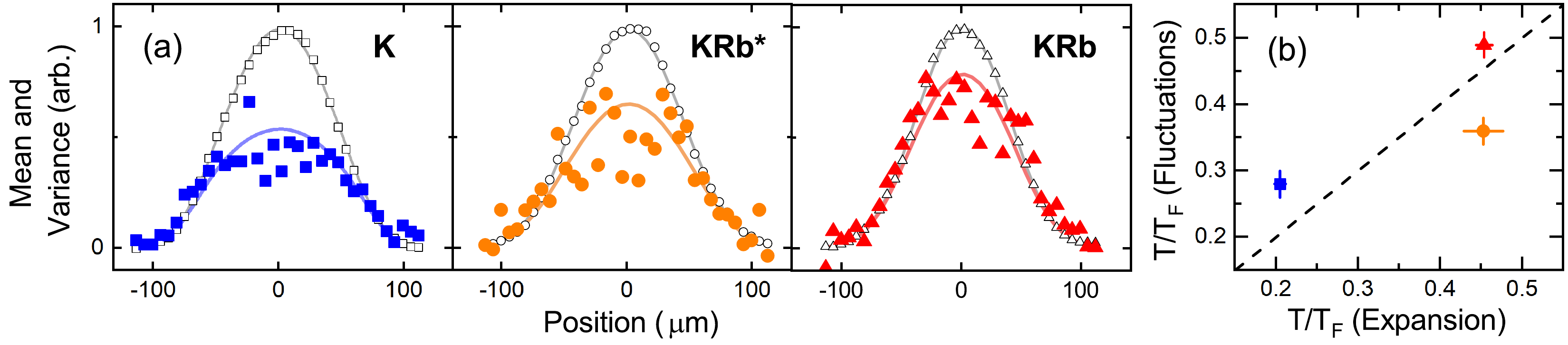}
\caption{(a) Mean (open symbols) and variance (solid symbols) profiles, in units of maximum particle number per bin, for K, KRb*, and KRb, averaged over the central 30 camera pixels of each image in the $z$-direction. Solid lines are fits of the mean and variance to the Fermi-Dirac distribution and Eq. (\ref{eq:suppression}), respectively, used to independently determine the $T/T_{\text{F}}$ of each set of images. (b) Comparison between $T/T_{\text{F}}$ extracted from both fitting methods for K (square), KRb* (circle) and KRb (triangle). Dashed line indicates equal $T/T_{\text{F}}$ between the two methods. 
In both panels, KRb variance is not corrected for STIRAP effects (see main text). Error bars are statistical and correspond to standard errors of the mean.}
\label{fig:ttf}
\end{figure*}
We post-select images to reduce shot-to-shot variation by automatically discarding outliers in total number and temperature, and do not manually exclude any images. Final analysis is performed on 50--60 absorption images of each species. We subdivide the images into bins and fit each to the Fermi--Dirac momentum distribution. Subtracting each fitted profile from the raw optical density profile normalizes against total particle number fluctuations, which would otherwise be the dominant contribution to the variance. By additionally subtracting technical sources of variance---photon shot noise, camera readout noise, and saturation corrections---we obtain the particle number mean and variance for each bin in the set of images. The measured variance is scaled up to account for imaging resolution and depth-of-field effects, using a scaling factor of 2.2 determined from experimental measurements and simulations of non-degenerate K \cite{SI}. Finally, we compute $\sigma_N^2/\overline{N}$ for each bin across all images and compare it to theoretical predictions in order to extract the average $T/T_{\text{F}}$ of the set of images.

Figure \ref{fig:krb} shows the dependence of number variance on mean number for degenerate KRb and non-degenerate K. In the non-degenerate case, the variance has linear scaling with mean number over the entire gas, the result expected from Poissonian statistics. By contrast, the degenerate KRb exhibits non-linear scaling of fluctuations, which separate into two distinct regimes. At the edge of the cloud, corresponding to bins with the lowest mean molecule number, the fluctuations are Poissonian due to high availability of unfilled states near the Fermi surface. At the center of the cloud, corresponding to bins with the highest mean molecule number, the fluctuations are sub-Poissonian since most states are filled.

Integrating the particle density in the imaging direction and using the local density approximation, the spatial profile of the variance suppression is given by
\begin{equation}
\frac{\sigma_N^2}{\overline{N}}=\frac{\text{Li}_{1} (-\zeta e^{-V(x,z)/k_BT})}{ \text{Li}_{2} (-\zeta e^{-V(x,z)/k_BT})}, \label{eq:suppression}
\end{equation}
where $\zeta$ is the peak fugacity, $V(x,z)$ is the optical potential in the imaging plane (scaled by the TOF), and $\text{Li}_i$ is the polylogarithm function of order $i$ \cite{Muller2010}. Since here $V(x,z)$ is harmonic, \textit{in situ} number fluctuations are preserved in TOF \cite{Sanner2010}. Figure \ref{fig:ttf}(a) shows profiles of the mean and variance for each species, obtained in separate experimental runs. The suppression is largest at the center of the gas and is reduced approaching the edges, due to the spatial profile of the trapping potential. Independent measures of $T/T_{\text{F}}$ are obtained by (i) fitting the cloud profile in expansion or (ii) fitting the variance suppression to Eq. (\ref{eq:suppression}) and extracting $\zeta$. Comparing the $T/T_{\text{F}}$ fit from both methods, we find close agreement for all species across a large range of $T/T_{\text{F}}$ (Fig. \ref{fig:ttf}(b)).

STIRAP transfers KRb* to KRb with a measured efficiency of 85\%, producing a slightly out-of-equilibrium initial KRb distribution and increasing the observed number fluctuations.
In the general case of molecule formation in bialkali atomic mixtures, STIRAP efficiency poses a technical limitation on degeneracy in the absence of ground-state molecule thermalization \cite{SI}.
For KRb* with $T/T_{\text{F}}=0.4$, the occupation fraction of the lowest-energy state in the trap is approximately 0.77. After the application of STIRAP, treated as a binomial process with uniform conversion efficiency over the entire molecular distribution, the occupation fraction is reduced by 15\%. The resulting increase in number fluctuations corresponds to a 17\% increase in $T/T_{\text{F}}$ \cite{SI}. For these conditions, the effect of STIRAP is small since the number of thermal holes is still significant.

When KRb is transferred back to KRb* for imaging, the second STIRAP sequence introduces additional holes; however, since the physical KRb distribution is not affected, this is an imaging artifact and can be corrected \cite{SI}. Accounting for the added variance reduces the KRb $T/T_{\text{F}}$ extracted from Eq. (\ref{eq:suppression}) from 0.49(2) to 0.44(2). As shown from the comparison to expansion profile fitting in Fig. \ref{fig:ttf}(b), the overall effect of STIRAP on the state occupation is minimal.

We have shown that atom--dimer elastic collision processes thermalize Feshbach molecules during conversion and enable the production of a degenerate molecular Fermi gas at equilibrium. Sub-Poissonian density fluctuations measured in degenerate Feshbach and ground-state molecules provide independent thermometry and a consistent picture of thermal equilibrium. In future experiments on polar molecular gases, local measurement of fluctuations could be used as a sensitive probe of many-body correlations.

We are grateful to E. Cornell, J. D'Incao, P. He, K.-K. Ni, A.M. Rey, and in particular W. Ketterle and C. Sanner, for stimulating discussions. This work was supported by DARPA DRINQS, ARO-MURI, AFOSR-MURI, NIST, and NSF PHY-1734006.

${}^*$W.G.T. and K.M. contributed equally to this work.

\bibliography{notes.bib}
\end{document}


\raggedbottom

\title{Supplementary Materials}

\maketitle

\section{Ultracold Bose--Fermi mixture}

The experimental sequence for preparing the K--Rb mixture has been described in \cite{DeMarco2019}. In Ref. \cite{DeMarco2019}, the optical traps crossed at an angle of $45^\circ$; in the current work, they cross at $90^\circ$. The trap frequencies are $(\omega_{x}, \omega_{y},\omega_{z}) = 2\pi \times (60, 240, 60)$ Hz for K, and are scaled by factors of 0.72, 0.83, and 0.79 for Rb, KRb*, and KRb, respectively. The trap $y$-axis and the bias magnetic field are aligned in the direction of gravity. For the measurement of $a_{\text{ad}}$, absorption imaging is performed along the trap $x$-axis. For the measurement of fluctuation suppression, absorption imaging is performed along the trap $y$-axis.

\section{Collisional damping}
The coupled center-of-mass oscillations of K and KRb* are described by
\begin{align}
\label{eq:damping_full}
\ddot y_{\text{K}} &= - \omega_{\text{K}}^2 y_{\text{K}} - \frac{4}{3} \frac{m_{\text{KRb}}}{M} \frac{N_{\text{KRb*}}}{N} \Gamma (\dot y_{\text{K}} - \dot y_{\text{KRb*}}) \\ 
\ddot y_{\text{KRb*}} &= - \omega_{\text{KRb*}}^2 y_{\text{KRb*}} + \frac{4}{3} \frac{m_{\text{K}}}{M} \frac{N_{\text{K}}}{N} \Gamma (\dot y_{\text{K}} - \dot y_{\text{KRb*}}), \nonumber
\end{align}
where $y$ is the displacement from the equilibrium position, $M = m_{\text{K}} + m_{\text{KRb}}$, $N=N_{\text{K}} + N_{\text{KRb*}}$, $\Gamma = \overline n \sigma v_{\text{rel}}$ is the K--KRb* collision rate, and $\omega_{\text{K}}$ ($\omega_{\text{KRb*}}$) is the trap frequency in the y-direction for K (KRb*) \cite{Ferrari2002}. The overlap density $\overline n$ between K and KRb* is
\begin{align}
\overline n &= \left(\frac{1}{N_{\text{K}}} + \frac{1}{N_{\text{KRb*}}} \right)\; \int n_{\text{K}} n_{\text{KRb*}} \; d^3x  \\
	&= N\left(\frac{2\pi k_{\text{B}} T_{\text{K}} }{m_{\text{K}}} \right)^{-3/2}  \overline{\omega}_{\text{K}}^3 \left(1 + \frac{m_{\text{K}}} {m_{\text{KRb}}} \frac{T_{\text{KRb*}}}{T_{\text{K}}} \frac{1}{\gamma^2}  \right)^{1/2}, \nonumber
\end{align}
where $\overline \omega_{\text{K}}$ is the geometric mean trap frequency for K, and $\gamma = 0.83$ is the ratio between the trap frequencies for K and KRb*. Here we have assumed Boltzmann distributions for both species, since $T/T_{\text{F}} \approx 1$ for both K and KRb* for the measurement of $a_{\text{ad}}$. The thermally-averaged relative velocity is
\begin{equation}
v_{\text{rel}} = \sqrt{\frac{8 k_{\text{B}}}{\pi} \left(\frac{T_{\text{K}}}{m_{\text{K}}} + \frac{T_{\text{KRb*}}}{m_{\text{KRb}}} \right)}.
\end{equation}

For the experimental number ratio ($N_{\text{K}} / N_{\text{KRb*}} \approx 7.5$) and collision rate, and for initially stationary K, the motion of the two species (Eq. \ref{eq:damping_full}) is nearly uncoupled, and we obtain
\begin{equation}
\ddot y_{\text{KRb*}} \approx - \omega_{\text{KRb*}}^2 y_{\text{KRb*}} - \frac{4}{3} \frac{m_{\text{K}}}{M} \frac{N_{\text{K}}}{N} \Gamma \dot y_{\text{KRb*}},
\end{equation}
which gives an exponential damping time $\tau$ of
\begin{equation}
\frac{1}{\tau} = \frac{2}{3} \frac{m_{\text{K}}}{M} \frac{N_{\text{K}}}{N} \Gamma. \label{eq:damping_approx}
\end{equation}
No induced oscillation of K is observed in the experiment, which supports the use of this approximation. The damping rate from the approximate expression (Eq. (\ref{eq:damping_approx})) differs from the full solution (Eq. (\ref{eq:damping_full})) by less than $3\%$ over the full range of our experimental parameters. This error is much smaller than the uncertainty on the experimentally-measured damping time.

In the experiment, we measure the exponential damping rate $1/\tau_1$ of KRb* in the presence of K. There is additionally a small damping $1/\tau_0$ due to the anharmonicity of the trap, which we measure by preparing a gas of KRb* without K and exciting the same amplitude of oscillation. Hence, the damping due to elastic collisions with K is $1/\tau_{\text{el}} = 1/\tau_1 - 1/\tau_0$. Setting $\tau = \tau_{\text{el}}$ in Eq. (\ref{eq:damping_approx}) gives a measurement of the atom--dimer cross section $\sigma$.

\section{STIRAP Recoil-Induced Oscillations}

In Ref. \cite{DeMarco2019}, it was noted that the KRb molecules exhibited a large center-of-mass oscillation in the vertical direction after STIRAP, which was attributed to a differential gravitational sag between the atomic and molecular species. To prepare degenerate clouds, a weak vertical lattice was used to suppress this effect.

The two STIRAP beams also propagate along the vertical direction. Since the writing of Ref. \cite{DeMarco2019}, we learned that the KRb oscillation can be primarily attributed to the photon recoil of the STIRAP process given by $k_{\text{STIRAP}} = 2\pi/\lambda_1 - 2\pi/\lambda_2$, where $\lambda_1 = 690$ nm and $\lambda_2 = 970$ nm are the wavelengths of the two Raman lasers. Regardless of the physical origin, adding the weak vertical lattice effectively removes the oscillation.

\begin{figure}
    \centering
    \includegraphics[width=0.49\textwidth]{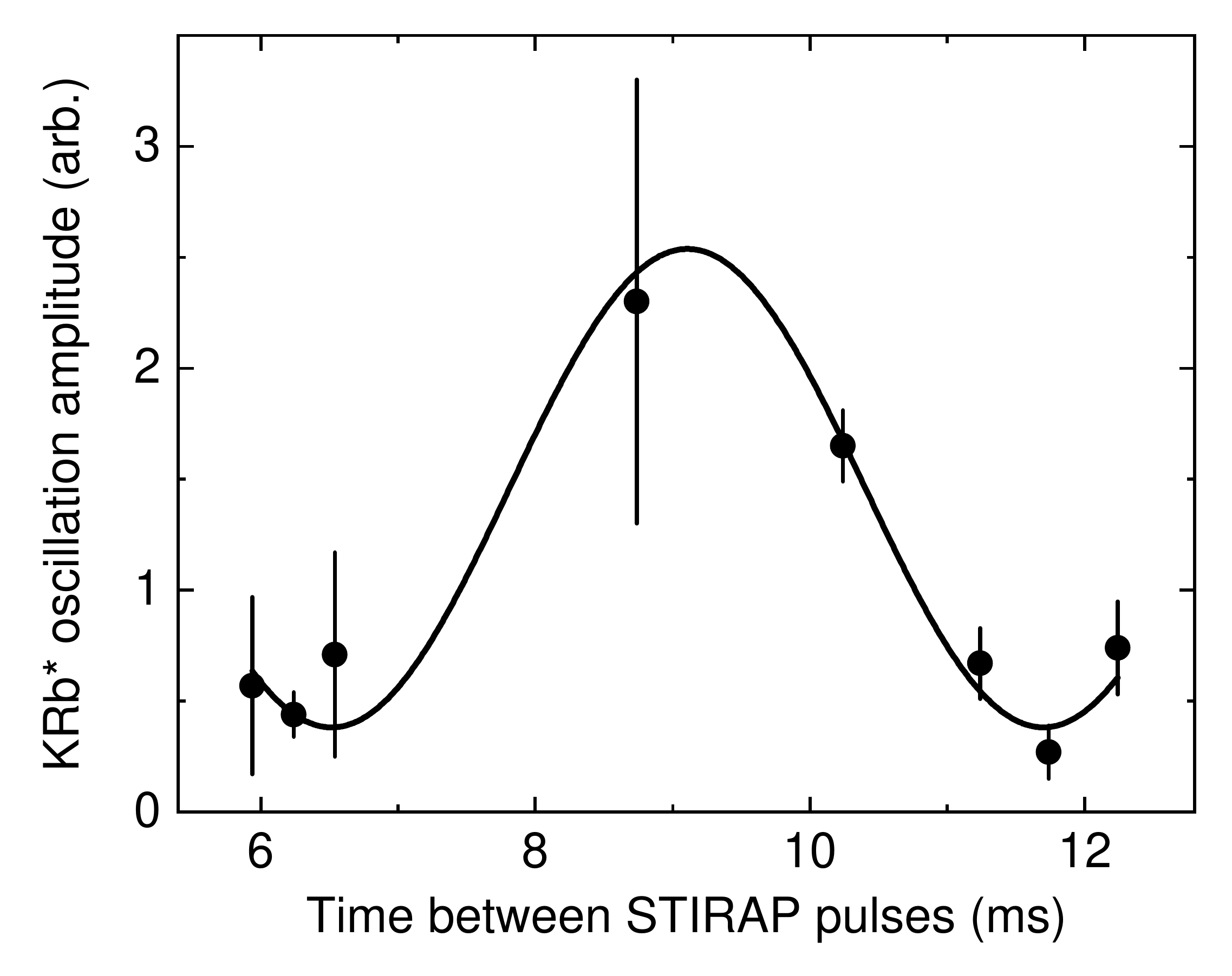}
    \caption{Amplitude of KRb* oscillations after two STIRAP pulses as a function of the delay time between the pulses. The solid line is a fit to a sinusoid with the frequency fixed at the measured trap frequency for the ground-state molecules ($\omega_{\text{y}} = 2\pi \times 190$ Hz). Oscillation damping (cross-species thermalization) measurements are performed with a timing where the oscillation amplitude is maximized (minimized).}
    \label{fig:stirap_timing}
\end{figure}

To measure $a_{\text{ad}}$, we use STIRAP to transfer KRb* to the ground state, remove most of the unpaired atoms, and use STIRAP again to transfer back to the Feshbach state. The weak vertical lattice is not used for this measurement. By changing the timing of the two STIRAP pulses relative to the vertical trap frequency $\omega_{\text{y}}/2\pi = 190$ Hz, the momentum recoils from the two pulses can be made to add up or cancel, giving control over the amplitude of the KRb* oscillation. If the STIRAP sequences are separated in time by one trap oscillation period, the pulses impart the maximum velocity of $v_{\text{STIRAP}} = 2 \hbar k_{\text{STIRAP}} / m_{\text{KRb}} = 2.6 \text{ mm/s}$, corresponding to a 15.5 $\mu$m oscillation amplitude after 6 ms TOF. This is in reasonable agreement with the experimentally-measured value of $\sim 21$ $\mu$m (shown in Fig. 1, upper panel of main text). Figure \ref{fig:stirap_timing} shows the amplitude of KRb* oscillations as a function of the delay time between the two STIRAP pulses, showing that the oscillation amplitude can be controlled using this timing.

\section{Cross-species thermalization}

In the preparation of the K--KRb* mixture for the oscillation damping experiments, the K temperature is initially much larger than the KRb* temperature. For the correct choice of delay time between the STIRAP pulses, KRb* does not oscillate after the pulses. However, due to the large atom--dimer elastic cross section, KRb* rapidly heats to the temperature of K. The timescale $\tau$ of this heating gives an alternative measurement of the elastic collision cross section, according to
\begin{equation}
    \tau = \frac{2.7/\xi}{\overline n \sigma v_{\text{rel}}},
\end{equation}
where 2.7 is the number of collisions to thermalize for $s$-wave collisions \cite{Monroe1993,Wu1996} and $\xi = 4 m_{\text{K}} m_{\text{KRb}} / (m_{\text{K}}+m_{\text{KRb}})^2 = 0.73$ quantifies the efficiency of thermalization between particles of different masses \cite{Mosk2001}.

\begin{figure}
    \centering
    \includegraphics[width=0.48\textwidth]{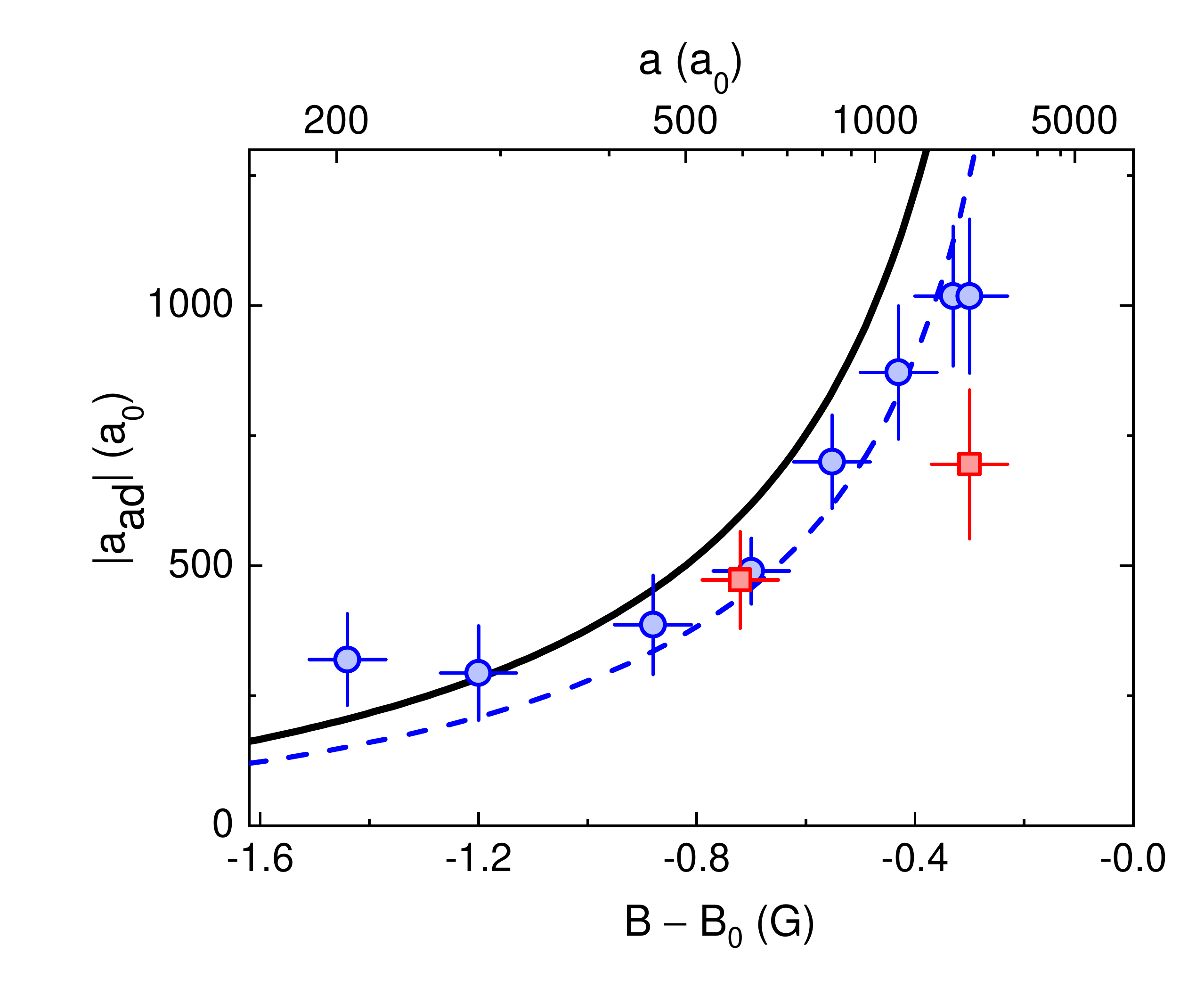}
    \caption{Measured K--KRb* scattering length $|a_{\text{ad}}|$ as a function of magnetic field. The results from cross-species thermalization measurements (squares) are in good agreement with the results from oscillation damping (circles, same data as Fig. 1, lower panel of main text). The K--Rb scattering length $a(B)$ (solid line) and the fit from the main text (dashed line) are shown for comparison.}
    \label{fig:cross-species}
\end{figure}

Figure \ref{fig:cross-species} shows the $|a_{\text{ad}}|$ extracted from cross-species thermalization measurements. For comparison, the results are plotted with the $|a_{\text{ad}}|$ measured with collisional damping (same data as lower panel of Fig. 1 in the main text). The two methods show reasonable agreement. Collisional damping was chosen as the primary technique for measuring $|a_{\text{ad}}|$ since the signal-to-noise was better than that of the cross-species thermalization measurements.

\section{\NoCaseChange{Rb--KRb*} elastic collisions}

We also measured damping of KRb* dipole oscillations through elastic collisions with Rb. Since Rb is the minority species, the Rb number after making Feshbach molecules is more unstable than the K number, and the measurement suffered from a large uncertainty of the Rb number. This affects the uncertainty of the extracted $a_{\text{Rb--KRb*}}$ through the uncertainty of the overlap density $\overline n$, and introduces a systematic uncertainty since the Rb and KRb* oscillations become more coupled as $N_{\text{Rb}}$ and $N_{\text{KRb*}}$ become comparable ($N_{\text{Rb}}/N_{\text{KRb*}}$ varied from 0.25 to 1 for this measurement).

\begin{figure}
    \centering
    \includegraphics[width=0.48\textwidth]{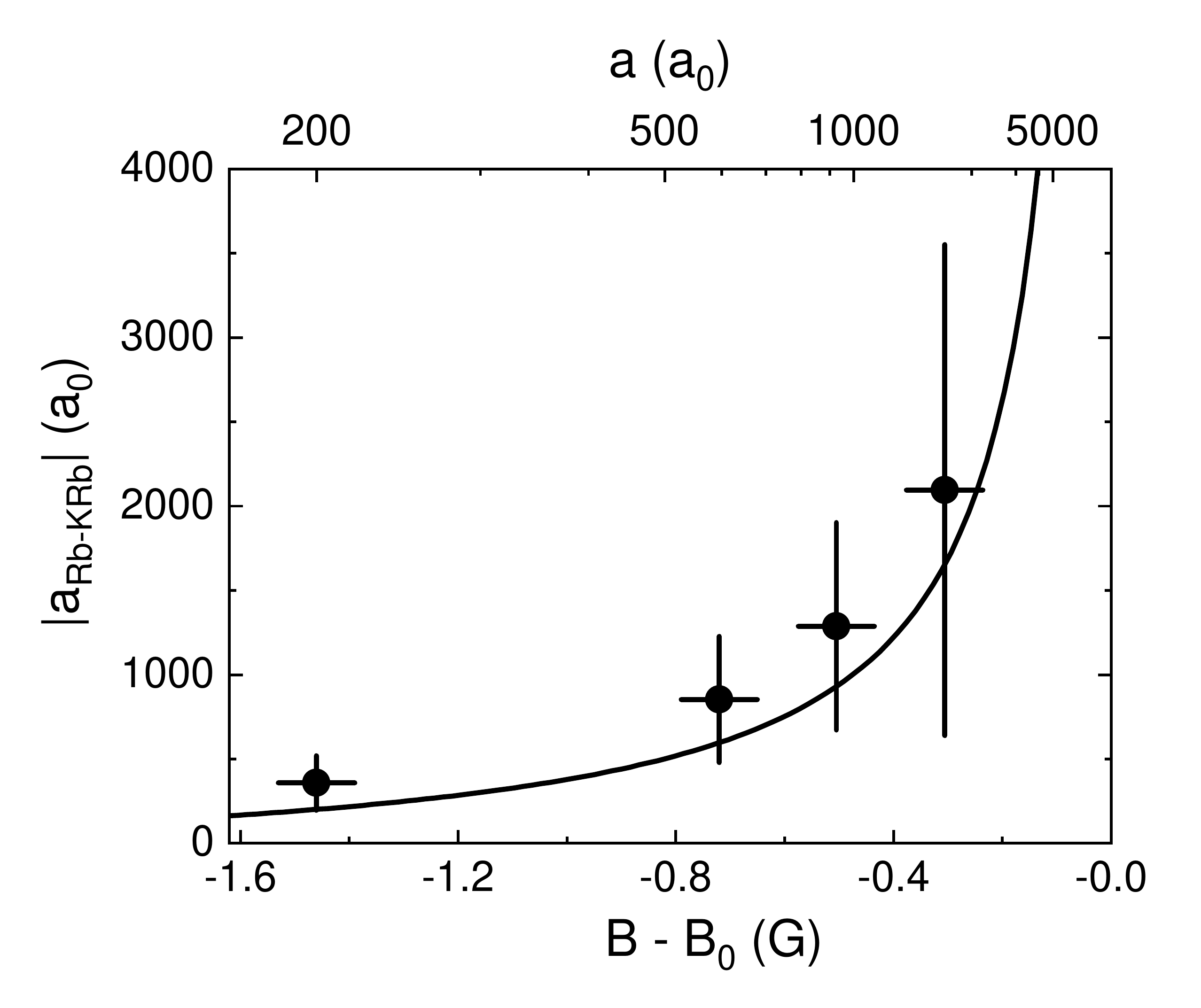}
    \caption{Measured $|a_{\text{Rb--KRb*}}|$ from oscillation damping measurements, with the K--Rb scattering length $a(B)$ (line) shown for comparison. The extracted $|a_{\text{Rb--KRb*}}|$ shows an upward trend as the resonance is approached.}
    \label{fig:aRbKRb}
\end{figure}

Figure \ref{fig:aRbKRb} shows the extracted Rb--KRb* scattering length $|a_{\text{Rb--KRb*}}|$ as a function of magnetic field. The measurements were performed with an overlap density of $\overline{n} = 1.2(6) \times 10^{12} \text{ cm}^{-3}$, and at a temperature of $T = 200$ nK. The large error bars are dominated by the uncertainty of the Rb density.

\section{Number of elastic collisions}

In the main text, we estimate that the Feshbach molecules experience at least 6 elastic collisions during the Feshbach ramp. Since the Feshbach ramp is a linear ramp, we calculate the magnetic field-averaged elastic K--KRb* collision cross section,
\begin{equation}
    \overline{\sigma} = \frac{1}{\Delta B} \int_{B_0-\Delta B}^{B_0} dB\, \left(\frac{4\pi a_{\text{ad}}^2}{1 + k_{\text{th}}^2 a_{\text{ad}}^2}\right) = 3.2 \times 10^{-10} \text{ cm}^{2},\label{eq:avg_sigma}
\end{equation}
where $\Delta B = 546.62 \text{ G} - 545.5 \text{ G} = 1.12$ G and $a_{\text{ad}}(B) = c a(B)$.
The upper bound $B_0$ of the integral is the Feshbach resonance position, and the lower bound is the endpoint of the Feshbach ramp. Above, $1/k_{\text{th}} = \sqrt{\hbar^2 / 2\mu k_B T_{\text{F}}^{\text{K}}} = 2000a_0$, $T_{\text{F}}^{\text{K}} = 680$ nK is the Fermi temperature for K, and $c = 0.74$ is the best-fit value from the main text.

The number of elastic collisions is given by
\begin{equation}
    N_{\text{el}} = \overline{n}\, \overline{\sigma} v_{\text{rel}} \times \Delta t,
\end{equation}
where $\overline n$ is the overlap density, $v_{\text{rel}} = \sqrt{\frac{8k_B}{\pi}\left(\frac{T_{\text{F}}^{\text{K}}}{m_{\text{K}}} + \frac{T_{\text{F}}^{\text{KRb*}}}{m_{\text{KRb}}}\right)}$ is the average relative velocity, $T_{\text{F}}^{\text{KRb*}} = 220$ nK is the Fermi temperature for KRb*, and $\Delta t$ is the time spent on the molecular side of the Feshbach resonance during the ramp. We assume a cloud of $5 \times 10^5$ K atoms at $T/T_{\text{F}}^{\text{K}} = 0.1$ and $3 \times 10^4$ KRb* molecules at $T/T_{\text{F}}^{\text{KRb*}} = 0.3$, both at $T = 70$ nK, which gives an overlap density of $\overline n = 1.6 \times 10^{13} \text{ cm}^{-3}$. For a typical 5 ms ramp from 555 G to 545.5 G, $\Delta t = 0.6$ ms, and the estimated number of elastic collisions is $N_{\text{el}} = 5.9$. If the average energy per particle $\frac{3}{4} T_{\text{F}}$ is used in the calculations instead of $T_{\text{F}}$, we extract a slightly higher value of $N_{\text{el}} = 6.08$. This estimate does not include additional effects such as Pauli blocking of K collisions and the state distributions of K and KRb* when the molecules are formed. However, clear evidence of thermalization indicates that sufficiently many elastic collisions occur during the Feshbach ramp.

An upper bound on the number of inelastic collisions can be derived by assuming 100\% conversion of Rb to KRb*. Since the observed KRb* number at the end of the Feshbach ramp is $\sim 50\%$ of the initial Rb number \cite{DeMarco2019}, the maximum number of inelastic collisions per KRb* is $N_{\text{inel}} = 1$. However, a small amount of residual Rb is observed at the end of the Feshbach ramp, indicating $<100\%$ Feshbach conversion and suggesting that the true number of inelastic collisions is $N_{\text{inel}} < 1$ and that a small number of Rb--KRb* elastic collisions may contribute to thermalization.

\section{Fluctuation Measurement Conditions}

For the measurement of fluctuation suppression, we perform measurements on degenerate samples of K, KRb*, and KRb. For each species, a set of 100--125 absorption images are taken at 6 ms TOF. We use a low probe intensity $I/I_{\text{sat}} = 0.2$ to mitigate saturation effects, and correct the optical density for saturation of the atomic transition using Eq. (\ref{eq:od}). To reduce shot-to-shot variation, only images with total particle number within $\pm 15\%$ of the median number and with fitted fugacity greater than one ($T/T_{\text{F}}<0.57$) are retained, leaving between 50--60 images of each species for analysis.

In previous experiments on degenerate KRb, typical K conditions were $5 \times 10^5$ atoms at $T/T_{\text{F}} = 0.1$ \cite{DeMarco2019}. Such samples are too dense to measure particle number with the accuracy needed for this experiment. Therefore, the K conditions used for the fluctuation measurement are less degenerate than those used to create molecules. To create appropriate samples for the fluctuation measurement, we remove Rb atoms, prepare a spin mixture of K in the $|F, m_F\rangle = |9/2, -9/2\rangle$ and $|9/2, -7/2\rangle$ hyperfine states, and hold the mixture for three seconds to allow thermal equilibration. Varying the optical trap depth and the fraction of K in each spin state allows control over the final atom conditions. 

Typical K conditions for the measurement on degenerate atoms are $1 \times 10^5$ atoms at $T/T_{\text{F}} = 0.2$; for the measurement on non-degenerate atoms, typical conditions are $5 \times 10^4$ atoms at $T/T_{\text{F}} > 0.6$. Typical molecule conditions are $6 \times 10^4$ of KRb* and KRb at $T/T_{\text{F}}=0.45$.

\section{Number and Variance Counting}

To accurately measure  fluctuation suppression, it is essential to calibrate the conversion between particle number and optical density, and to characterize sources of optical density variance external to the actual particle number fluctuations. In absorption imaging, the optical density (OD) is given by
\begin{equation}
\label{eq:od}
    \text{OD}=\ln\left(\frac{P_l}{P_s}\right)+\frac{P_l-P_s}{P_{\text{sat}}^{\text{eff}}}
\end{equation}
where $P_s$($P_l$) is the difference in photon number between the shadow(light) and dark frames. $P_{\text{sat}}^{\text{eff}}=t I_{\text{sat}}^{\text{eff}}/(hc/\lambda)$ is the effective saturation intensity for the imaging system, in units of photon number, for fixed probe time $t$ and imaging wavelength $\lambda$. The conversion between optical density (OD) and the atom number $N$ on a single bin is $OD= \sigma_{\text{eff}} N/A$,
where $\sigma_{\text{eff}}$ is the effective imaging cross section and $A$ is the bin area. A single factor $\alpha$, quantifying the effects of imperfect polarization, imaging laser linewidth, and transmission losses in the imaging system, relates the effective quantities to their bare atomic counterparts:  $I_{\text{sat}}^{\text{eff}}=\alpha I_\text{sat}$ and $\sigma^{\text{eff}}=\sigma_0/\alpha$. To convert optical density measured in absorption imaging directly to atom number, given the details of the imaging system, the only missing parameter is $\alpha$ \cite{Hueck2017}.

We measure $\alpha$ by two complementary methods. First, by directly measuring the probe polarization, linewidth, and transmission, we find that the maximum physical absorption cross section is reduced by a factor $\alpha=1.66$. Second, we follow the procedure described in Ref. \cite{Reinaudi2007} to extract $\alpha$ by imaging the atoms: (i) we take a series of absorption images with fixed atom conditions, varying the probe intensity over a factor of ten between images; (ii) we extract the atom number from the OD using Eq. \ref{eq:od}, varying $\alpha$; and (iii) we choose the $\alpha$ that minimizes the variation of atom number with intensity. By this method, we extract $\alpha = 1.7(1)$, consistent with the direct measurement of the imaging light. For the measurements presented in the main text, we fix $\alpha=1.66$.

The variance of the optical density due to the probe light is
\begin{align}
    \sigma_{\text{probe}}^2&=\left(\frac{\partial \text{OD}}{\partial P_l}\right)^2 \sigma_{P_l}^2+
    \left(\frac{\partial \text{OD}}{\partial P_s}\right)^2 \sigma_{P_s}^2 \nonumber\\
    &=\left(\frac{1}{P_l}+\frac{1}{P_{\text{sat}}^{\text{eff}}}\right)^2\sigma_{P_l}^2+\left(\frac{1}{P_s}+\frac{1}{P_{\text{sat}}^{\text{eff}}}\right)^2\sigma_{P_s}^2
\end{align}
The camera has fixed readout noise variance per frame ($\sigma_r^2$), so the variance of each imaging frame is modified to become $\sigma_{P_{l/s}}^2\rightarrow\sigma_{P_{l/s}}^2+2\sigma_r^2$, where the factor of two accounts for the readout noise on both the light/shadow and dark frames. Using the fact that the imaging light has Poissonian variance (i.e. $\sigma_P^2/P=1$), this simplifies to
\begin{equation}
    \begin{split}
        \sigma_{\text{probe}}^2&=\frac{1}{P_l}+\frac{1}{P_s}+\frac{2\sigma_r^2}{P_l^2}+\frac{2\sigma_r^2}{P_s^2}+\frac{P_l+P_s}{(P_{\text{sat}}^{\text{eff}})^2}+\frac{4}{P_{\text{sat}}^{\text{eff}}}
    \end{split}
\end{equation}
By subtracting these quantities from the total measured optical density variance, the contribution from fluctuating atom number can be isolated:
\begin{equation}
    \sigma_N^2=\left(\frac{A}{\sigma_{\text{eff}}}\right)^2\left(\sigma_{\text{OD}}^2-\sigma_{\text{probe}}^2\right)
\end{equation}

\section{Finite Bin Size Effects}
In the measurement of number fluctuations, each image is divided into many smaller bins, corresponding to an integer number of CCD pixels. Since real imaging systems have finite resolution, the absorption signal from a single particle contributes to the signal measured on multiple adjacent bins. This effect tends to reduce the effective number variance on each bin, since holes in the particle distribution that fluctuate shot-to-shot are averaged across neighboring bins. A similar effect can occur due to particle motion during the imaging pulse; however, this effect is negligible for the probe time (10 $\mu$s) and particle mass in this experiment.

For bin sizes much larger than the imaging resolution, the correction becomes negligible. However, using larger bins reduces the total number of bins obtained from each image and makes it more difficult to accumulate statistics. The alternative route of improving the imaging resolution has the downside of reducing the depth-of-field, which broadens the absorption signal from atoms displaced from the focal plane. To balance these effects, we use a bin width of 6.45 $
\mu$m in the experiment, compared to the imaging resolution of approximately 2.5 $
\mu$m.

\begin{figure}
    \centering
    \includegraphics[scale=0.32]{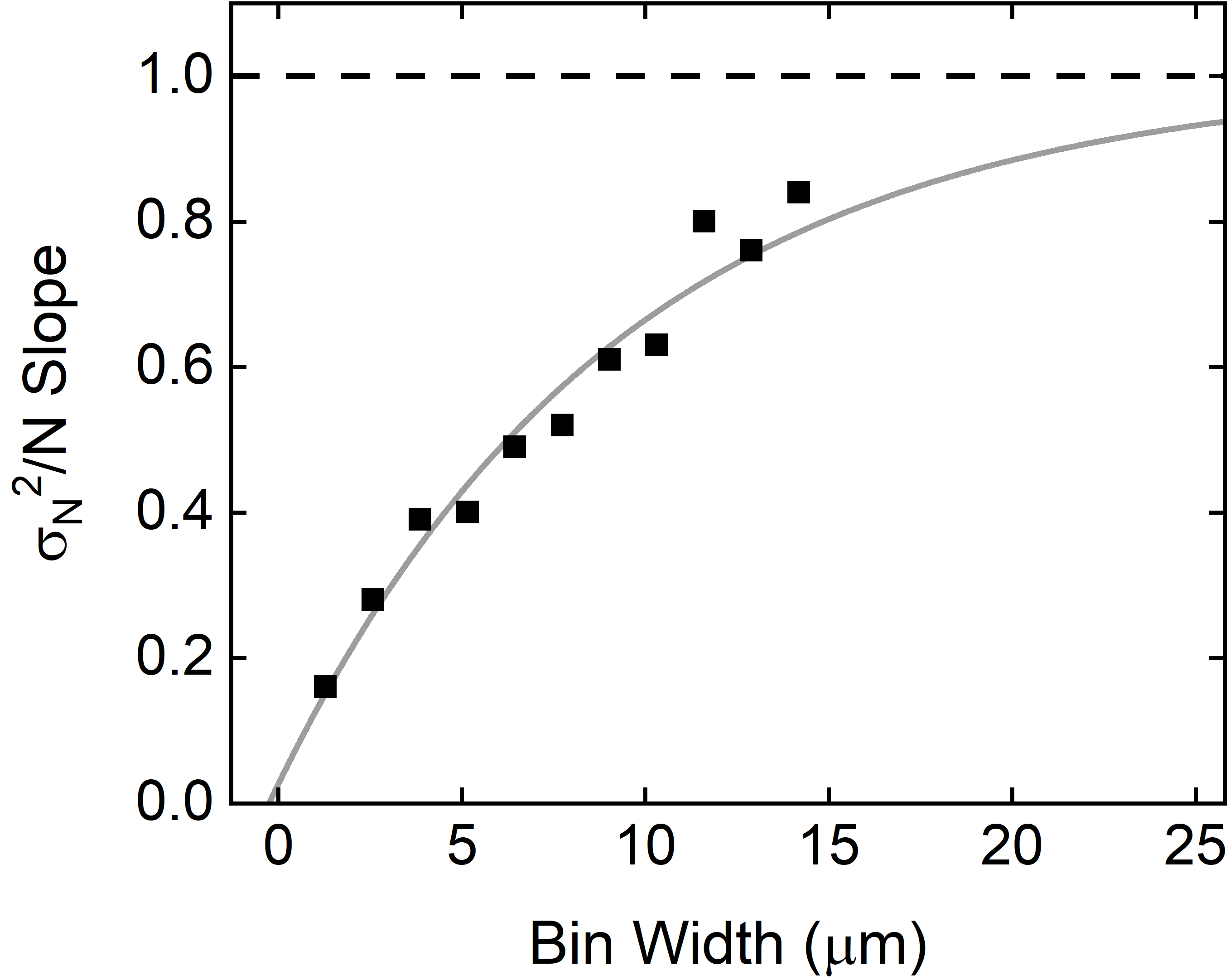}
    \caption{Variance vs. mean slope, as a function of bin width. Solid line is a guide to the eye. For increasing bin width, the slope approaches 1 (dashed line), the expected value for non-degenerate particles.}
    \label{fig:binsize}
\end{figure}
We quantify the scaling of the slope $\sigma_N^2/N$ by reanalyzing images of non-degenerate K ($T/T_\text{F}>0.6$) using bin widths ranging from 1.29--14.19 $
\mu$m (Fig. \ref{fig:binsize}). For a non-degenerate cloud, neglecting resolution effects, the expected slope is determined by Poissonian statistics:  $\sigma_N^2/N=1$. Over the range of bin widths, the slope is found to increase from 0.18 to 0.84, with the trend indicating saturation at the Poissonian value. The small size of the cloud prevents the use of larger bin widths for this measurement, but the slope saturating near 20-25 $\mu$m would be consistent with previous experiments that observed saturation at approximately 10 times the resolution \cite{Sanner2010,Amico2018}. For the bin width used in experiment, $\sigma_N^2/N$ extracted from experimental data was 0.45(4), and therefore all measured variance data was scaled up by a factor of 2.2 to recover the actual particle number variance. 

We additionally perform a simulation to verify the experimental results. Using the atom and molecule temperatures recorded in experiment, we generate simulated absorption images by random sampling from the Maxwell-Boltzmann momentum distribution. To approximate resolution and depth-of-field effects, we treat the absorption signal from each particle as a Gaussian beam, with the waist determined by the imaging resolution, and propagate the signal to the imaging plane centered on the cloud. For each simulated image, we bin and analyze the data equivalently to the experimental data, and average over many images. We extract a slope of 0.4 from the simulation, consistent with the experimental measurement.

\section{Effect of STIRAP on Occupancy}

The measurement of number fluctuations probes the occupation of states in the Fermi gas. The occupancy of the lowest-energy state in a Fermi gas is given by the Fermi-Dirac distribution
\begin{equation}
\label{eq: occ}
    f(\epsilon=0)=\frac{1}{e^{-\beta\mu}+1},
\end{equation}
where $f(\epsilon)$ is the occupation of states with energy $\epsilon$, $\beta=1/kT$ and $\mu$ is the chemical potential, and the product $\beta \mu$ is the logarithm of the peak fugacity $\zeta$. $\zeta$ is related to $T/T_{\text{F}}$ by
\begin{equation}
\label{eq: ttf}
    \frac{T}{T_{\text{F}}}=(6\,\text{Li}_3(-\zeta))^{-1/3},
\end{equation}
where $\text{Li}_i$ is the polylogarithm function of base $i$. When STIRAP is used to convert KRb* to KRb, it reduces the occupation of the gas since it does not convert KRb* with perfect efficiency.

Since KRb* is in thermal equilibrium after Feshbach association, it has a peak occupancy determined only by the peak fugacity: $f(\epsilon=0,\zeta)$. After STIRAP, with conversion probability $p$, the peak occupancy is modified to be $p f(0,\zeta)$, assuming no elastic collisions leading to rethermalization in the KRb gas. Even in the absence of thermalization, for high STIRAP efficiency KRb has a distribution near an equilibrium Fermi-Dirac distribution. We can relate the modified peak occupancy to a modified fugacity $\zeta'$ by making the equivalence
\begin{equation}
    p f(0,\zeta)=f(0,\zeta'),
\end{equation}
and then can use Eq. (\ref{eq: ttf}) to extract an effective $T/T_{\text{F}}$ after STIRAP.

\begin{figure}
    \centering
    \includegraphics[width=0.46\textwidth]{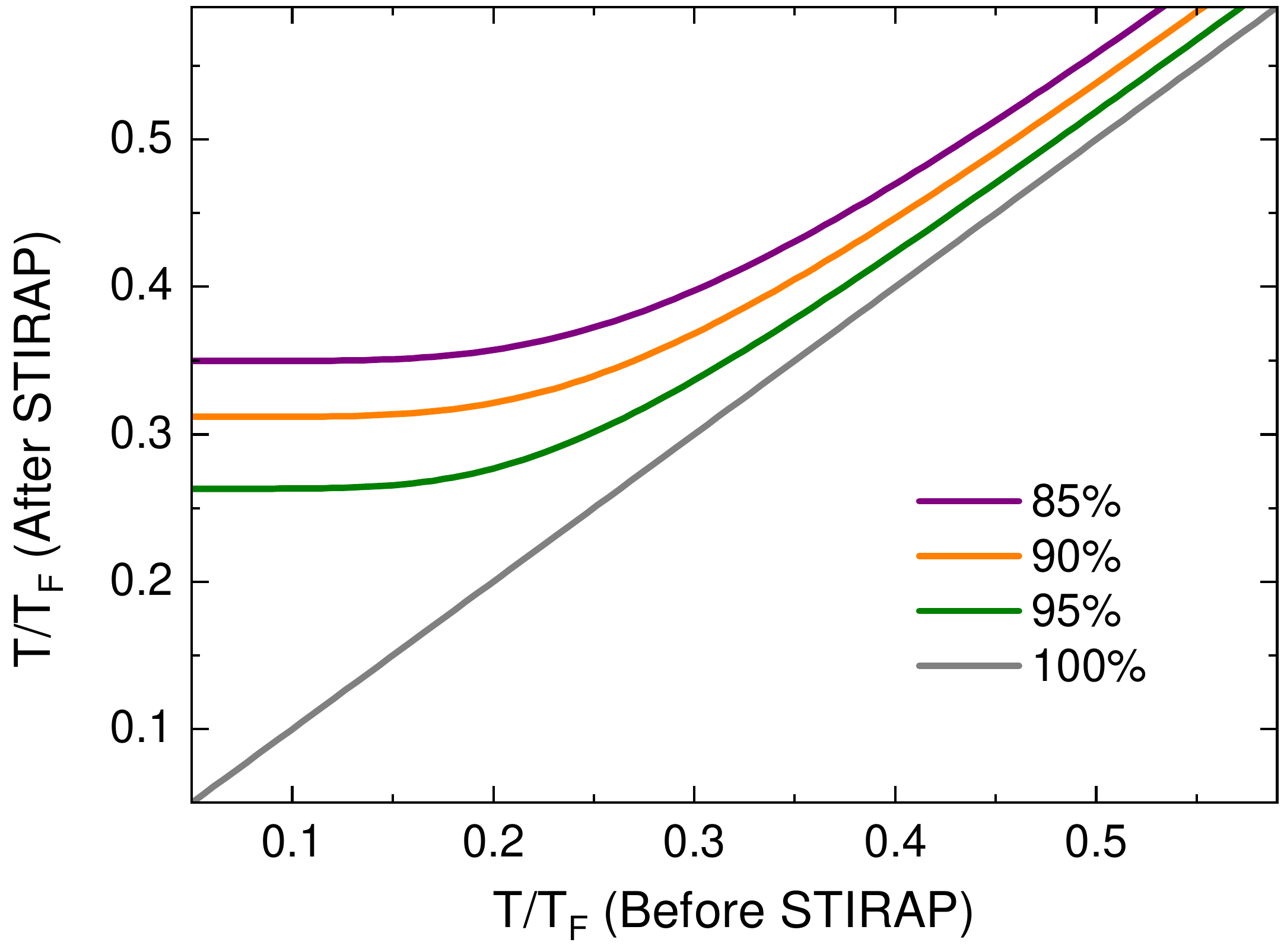}
    \caption{Modification of $T/T_{\text{F}}$ due to STIRAP with varying conversion efficiencies. The conversion efficiency in this experiment is $85\%$ (top curve, purple). }
    \label{fig:eff_ttf}
\end{figure}

Figure \ref{fig:eff_ttf} shows the effect on $T/T_{\text{F}}$ of STIRAP with conversion efficiencies ranging between $85\%$ and $100\%$.
The fractional effect of STIRAP on $T/T_{\text{F}}$ is smallest for high initial $T/T_{\text{F}}$, since the peak state occupation is initially low. For highly degenerate KRb* gases, by contrast, the peak occupation saturates at the STIRAP efficiency. To prepare a degenerate KRb gas with high peak occupancy, therefore, it is critical to have high STIRAP efficiency or to have thermalization processes occuring in the gas.

\section{STIRAP Imaging Variance}

Since STIRAP transfer is 85\% efficient, it introduces additional variance into the sample when KRb is converted to KRb* for imaging. This effect can be approximately quantified with a simple statistical model. The peak variance suppression in a subvolume of an atomic or molecular cloud has the form
\begin{equation}
\label{eq:polylog}
    \frac{\sigma_N^2}{\overline N}=\frac{\text{Li}_1(-\zeta)}{\text{Li}_2(-\zeta)}\equiv \eta,
\end{equation}
where $\sigma_N^2$ and $\overline N$ are the particle number variance and mean \cite{Muller2010}. 

After expansion, each subvolume comprises $N_k$ single-particle states. The probability of occupying each state is given by the Fermi-Dirac occupation number $f_k$. Since the occupation of each state is independent, the distribution of total particle number in the subvolume is the sum of independent binomial random variables over all $k$ states contained in the subvolume:
\begin{equation}
    N\sim \sum_{k}\text{Bin}(1,f_k),
\end{equation}
where Bin$(m,p)$ is the binomial distribution with $m$ trials, each with $p$ probability of success. Assuming that all of the $f_k$ are equal, which is valid in the limit of small subvolume size, this simplifies to
\begin{equation}
    N\sim \text{Bin}(N_k,f_k).
\end{equation}

In a subvolume with mean particle number $\overline N$ and variance $\sigma_N^2$, the values of $N_k, f_k$ can be written in terms of $\overline N,\sigma_N^2$ using the mean and variance of the binomial distribution:
\begin{align}
    \overline N&=N_k f_k \nonumber\\
    \sigma_N^2&=N_k f_k(1-f_k).
\end{align}

Substituting and simplifying according to Eq. (\ref{eq:polylog}):
\begin{align}
    N&\sim \text{Bin}(N_k,f_k) \nonumber\\
    &=\text{Bin}\left(\frac{\overline N^2}{\overline N-\sigma_N^2},\frac{\overline N-\sigma_N^2}{\overline N}\right) \nonumber\\
    &=\text{Bin}\left(\frac{\overline N}{ 1-\eta},1-\eta\right).
\end{align}

After applying STIRAP, modeled as a binomial process with probability $p=0.85$, the modified distribution is
\begin{equation}
    N'\sim \text{Bin}\left(\frac{\overline N}{ 1-\eta},p(1-\eta)\right).
\end{equation}
The modified ratio of mean and variance after STIRAP is therefore 
\begin{equation}
\label{eq:ratio}
    \frac{\sigma_{N'}^2}{\overline{N'}}=1-p(1-\eta).
\end{equation}

Using Eq. \ref{eq:ratio}, the modified fluctuations after STIRAP ($\sigma_{N'}^2/\overline{N'}$) can be used to extract $\eta$, which can then be used to calculate the peak fugacity via Eq. \ref{eq:polylog}. By accurately measuring the STIRAP efficiency, the effect of STIRAP in imaging can be subtracted.

\bibliography{notes.bib}